\documentclass[twocolumn]{article}
\usepackage[english]{babel}
\usepackage{amsthm}
\usepackage[utf8]{inputenc}
\usepackage{latexsym}
\usepackage{amsfonts}
\usepackage{amssymb}
\usepackage{amsmath}
\usepackage{amsthm}
\usepackage{graphicx}
\usepackage{hyperref}
\usepackage{booktabs} 
\usepackage{amsmath}
\usepackage{bbm}

\usepackage[margin=1.5cm, top=1.7cm, bottom=1.7cm]{geometry}

\theoremstyle{definition}
\newtheorem{definition}{Definition}[section]
\theoremstyle{theorem}
\newtheorem{theorem}{Theorem}[section]

\begin{document}
\title{{\bf A Random Walk based Trust Ranking in Distributed Systems} \vspace{1em}\\
\begin{large}
\begin{tabular}{ccc}
Alexander Stannat & & Johan Pouwelse \\
Delft University of Technology & & Delft University of Technology \\
A.W.Stannat@student.tudelft.nl & & peer2peer@gmail.com
\end{tabular}
\end{large}}
\date{\vspace{-5ex}}
\maketitle
\addtocontents{toc}{\setcounter{tocdepth}{-1}}
\section{Abstract}
{\bf
Honest cooperation among individuals in a network can be achieved in different ways. In online networks with some kind of central authority, such as Ebay, Airbnb, etc. honesty is achieved through a reputation system, which is maintained and secured by the central authority. These systems usually rely on review mechanisms, through which agents can evaluate the trustworthiness of their interaction partners. These reviews are stored centrally and are tamper-proof. In decentralized peer-to-peer networks, enforcing cooperation turns out to be more difficult. One way of approaching this problem is by observing cooperative biological communities in nature. One finds that cooperation among biological organisms is achieved through a mechanism called {\it {\bf indirect reciprocity}} \cite{Five Rules for the Evolution of Cooperation}. This mechansim for cooperation relies on some shared notion of \textit{trust}. In this work we aim to facilitate communal cooperation in a peer-to-peer file sharing network called Tribler, by introducing a mechanism for evaluating the trustworthiness of agents. We determine a trust ranking of all nodes in the network based on the Monte Carlo algorithm estimating the values of Google's personalized PageRank vector. We go on to evaluate the algorithm's resistance to sybil attacks, whereby our aim is for sybils to be assigned low trust scores.  
}

\section{Introduction}
\label{sec:Introduction}
\subsection{Historical Perspective}
\label{subsec:HistoricalPerspective}
With the ever-growing expansion and widespread acceptance of the internet, the field of research in distributed systems is gaining evermore importance. In its simplest definition, a distributed system is a group of different processors working together on a common task. These processors have a shared state, operate concurrently and can fail independently. The primary advantages of a distributed system over a centralized system are scalability, fault-tolerance and lower latency. There are however some drawbacks to the decentralized nature of these networks. The most notable being resource-management.\vspace{1em}\\

\noindent A particular example of distributed systems are peer-to-peer networks, also known as P2P networks. A peer-to-peer network allows computers to communicate without the need for a central server. Peer-to-peer file sharing refers to the distribution of digital media over a P2P network, in which the files are located on individuals' computers and shared with other members of the network. P2P software was the piracy method of choice in the early 2000s with software programs such as LimeWire, Gnutella and the BitTorrent client being the most prominent applications \cite{The Early Days of Mass Internet Piracy Were Awesome Yet Awful}. A Supreme Court decision in 2005 led to the closure of many of these sites for illegally sharing copyrighted material, mostly music.\vspace{1em}\\

\noindent In a peer-to-peer file sharing network agents up- and download files over the network to one another in a cooperative manner, whereby an agent that is holding a file (or at least a part of it) can offer it to other agents that require it, through actions called seeding and leeching. Seeders are those who offer upload bandwidth while leechers are the agents that download the data. While these types of systems have many advantages over the standard client-server model, they do have one fundamental problem: users have an obvious incentive to download, i.e. to leech, but no inherent incentive to seed. This results in behavior called lazy freeriding. \vspace{1em}\\

\noindent In order to avert freeriding and to incentivize its users to participate in the network reciprocally, early file sharing networks such as the BitTorrent protocol employ a mechanism called tit-for-tat \cite{Peer-to-peer networking with BitTorrent}. Tit-for-tat is a highly effective strategy in game theory for the iterated prisoner's dilemma, in which an agent cooperates first and then replicates it's contender's previous actions. In practice, this works as follows. Peers in the BitTorrent network have a limited number of upload slots to allocate. An agent will begin by exchanging upload bandwidth for download bandwidth with a number of its peers. If one of these peers turns out to be a leecher, i.e. does not reciprocate, it will be choked. This means the agent will discontinue it's cooperation and assign the corresponding upload slot to another randomly chosen peer in a procedure known as optimistic unchoking. \vspace{1em}\\

\noindent However, we find that defecting is the dominant strategy in the Prisoner's dilemma \cite{An optimal strategy to solve the 
Prisoner's Dilemma}. The agents' inability to coordinate and build expectations of their counterparts ensures that everyone will be worse of than if they had collaborated. This is also known as the tragedy of the commons. Agents do not keep a memory about their peer's reliability which enables such lazy freeriding and other types of uncooperative behavior. There is no mechanism with which peers can be evaluated based on their previous reliability or trustworthiness as nodes in the network and hence every new transaction entails the risk of the partner node defecting. This leads us to the main research question, the TU Delft's Blockchain Lab focuses their study on: Is it possible to incorporate a digital counterpart to \textit{trust} in a distributed network with no central authority?

\subsection{Trust}
\label{subsec:Trust}
Trust is a rather abstract concept which is oftentimes understood more on an intuitive level as opposed to having a clear-cut definition. Thus there are many different possible definitions of trust. In this project we chose to adopt one of the definitions of trust given in Vandenheuvel's {\it Mathematics of Trust} \cite{Mathematics of Trust}.\\

\begin{definition}[Trust]\ \\
Two agents (a trustor and a trustee) have a trust relationship if the trustor is uncertain about the outcome of the trustee's actions and can only rely on previously developed expectations to predict these.
\end{definition}

\noindent Trust is a necessary catalyst for cooperation in networks and there have been many different approaches to facilitating trust relationships in distributed systems. Companies like eBay, Uber or Airbnb utilize reputation systems based on ratings. Agents in these networks leave publicly accessible reviews of eachothers' collaborators after each transaction. These reviews are stored and maintained by a central authority, so that no fraud can take place. Based on these reviews the network generates reputation systems for all nodes. Agents then decide who to interact with based on this reputation, thereby incentivizing cooperative behavior of all participants \cite{Reputation Systems}. \vspace{1em}\\

\noindent The TU Delft's Blockchain Lab aims to take this trust system a step further, by removing the central authority that manages the network and its reputation system. In order to gain an understanding of how this is possible we look into the sociological aspects of cooperative networks among people in the real world. Nowak, Martin A. discuss what psychological mechanisms engender cooperative networks in the animal kingdom, namely kin selection, direct reciprocity, indirect reciprocity, network reciprocity, and group selection \cite{Five Rules for the Evolution of Cooperation}. DBL is researching the possibility of incorporating a mechanism of indirect reciprocity into their p2p file sharing network called tribler \cite{TRIBLER:a social-based peer-to-peer system}.

\subsection{Trustchain}
\label{subsec:Trustchain}
In order to build such a digital trust mechanism into a perfectly decentralized network, a distributed storage space, or ledger, is required. The tools most commonly used for this purpose are Blockchains. Blockchains are data structures that utilize cryptographic primitives such as public-key cryptography to maintain a consensus on data, stored on many different processors in a distributed system. Transactions between agents in the network are grouped in blocks which, in turn, are interlinked by a hash chain. Blocks are created by "miners", nodes in the network that collect and group transactions. In order to obtain a block, the miner needs to solve a crpytographic hash puzzle through a protocol known as proof-of-work. If conflicting states occur, the chain forks, and miners contribute to the chain they believe is the valid one. At some point, one chain will overtake the other and all miners transition to {\it that} chain. This point is determined by a certain number of blocks by which one chain surpasses the other, which is based on a predetermined lower bound for the probability of a dishonest miner single-handedly overruling the current chain. The resulting chain of blocks is therefore immutable as well as fraud-proof\cite{Bitcoin: A peer-to-peer electronic cash system}. \vspace{1em}\\

\noindent Blockchains however have a major drawback that the classical client-server model does not have. Consensus is maintained by miners receiving information about all transactions that have transpired in the network. As the network grows, the risk of transaction broadcasts not reaching certain miners increases, which makes maintaining a global consensus more and more difficult. This fundamentally limits the scalability of such blockchain based networks. Another scalability issue the current proof-of-work consensus mechanism causes, lies in the fact that it requires agents to wait for a certain number of blocks to exceed a transaction's block before this transaction is deemed trustworthy. In pursuing a more scalable alternative, the blockchain lab has developed their own type of distributed ledger, called TrustChain. TrustChain is what's known as a fourth-generation blockchain. \vspace{1em}\\

\begin{figure}
\includegraphics[scale=0.6]{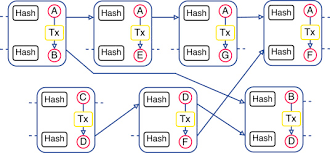}
\caption{Trustchains of different network participants}
\label{fig:Trustchain}
\end{figure}

\noindent{} In TrustChain, all network participants maintain their own chain of transactions. There is no mining and no global consensus. The TrustChain maintains records of all interactions between peers in the network, in respective blocks. Each block contains data about an individual transaction between two peers, such as the respective up-and download values, the peers' public keys and signatures as well as block sequence numbers and respective hashes. Blocks are linked to one another through hash pointers. Each block is thereby connected to two preceding and two succeeding blocks, i.e. each block is contained in the chains of both transaction partners. This results in many different chains, each corresponding to a single agent's transaction history, see Figure \ref{fig:Trustchain}\cite{TrustChain: A Sybil-resistant scalable blockchain}. \vspace{1em}\\

\noindent When two agents interact with one another, they make their respective chains visible to each other and may even store information about each other's chains as well. This structure is strongly scalable, both in the number of agents in the network as well as in the number of transactions per agent. Moreover, the trustchain does not maintain a global consensus. This means that double-spends are not actually prevented, as they are in traditional blockchains. However, they are made detectable and can subsequently be penalized through peer rejection or even by banning dishonest nodes from the network. Thereby fraudulent activity is not actually prevented, but strongly disincentivized.

\subsection{Problem Description}
\label{subsec:ProblemDescription}
As discussed in subsection \ref{subsec:HistoricalPerspective} (Historical Perspective) the long standing issue of tribler is digitizing a method for evaluating the trustworthiness, or reputation, of agents in a network. Such a mechanism is meant to deter lazy freeriders, i.e. agents that purposefully contribute no or very little resources to the network, but at the same time utilize the network for their own benefit. \vspace{1em}\\

\noindent There have been many different approaches to this problem, however a viable and accurate solution has yet to be found. The first reputation system introduced in Tribler was BarterCast \cite{Bartercast: A practical approach to prevent lazy freeriding in p2p networks}. BarterCast was based on the voluntary reports of agents about their own transactions. However, an inherent problem with the BarterCast system is the issue of deliberate misreporting of transactions, by agents that have made overall negative contributions to the network. After BarterCast, another accounting mechanism was introduced: DropEdge, which is an implementation of BarterCast on a subset of the graph, ignoring all nodes that are interested in receiving some data. \vspace{1em}\\

\noindent In this project we aim to implement a mechanism which enables the ranking of nodes in the tribler network based on their respective reputability. Reputatbility should be a measure of the impact a node has on the network. This means that nodes with a low or negative net contribution made to the network should be ranked lower than more altruistic nodes. When a node is looking to query an overlay for a file to download, it will generate this ranking and determine which agent in the swarm it should engage with to maximize the likelihood of a successful transaction and a subsequent reciprocal relationship. We present a number of requirements that this trust mechanism should satisfy.

\begin{itemize}
\item[1] {\bf Personalization} \\ Trust is personal. Two different agents in the network may have a completely different perception of their neighbours' trustworthiness. This means that an algorithm computing a trust ranking of nodes must depend on the node from whose perspective the trustworthiness is determined.
\item[2] {\bf Locality} \\ It's been observed that in the BitTorrent network most peers share a one hop relationship In \cite{One Hop Reputations for Peer to Peer File Sharing Workloads}. Therefore we restrict the amount of indirection in between contributing and reciprocating peers. This makes sense in a social context as well. A node may trust one of its peers, and by transitivity of trust, it will also trust this peer's trustees, although less. A third or fourth hop will reduce this amount of transitive trust to a minimum.
\item[3] {\bf Incrementality} \\
The underlying graph structure of the network continually evolves as transactions take place and new blocks are added to agents' trustchains. Trust rankings therefore develop continuously as well. Recomputing these trust rankings from scratch is prohibitively expensive. We require our algorithm to be able to update the trust rankings incrementally as more information arrives over time. We update the trust rankings in batches, either in time intervals, or by the amount of information that has become available. \vspace{1em}\\
\end{itemize}

\noindent A very popular existing algorithm that is used to rank the importance of pages on the world wide web is Google's pagerank algorithm. In the context of the internet, we find that the importance ranking by google is an equivalent concept to our idea of trustworthiness in p2p networks. Given the resemblance of the graph structures of both the web and the tribler network, the use of the pagerank algorithm for the assessment of agents' trustworthiness suggests itself.

\section{Preliminaries}
\label{sec:Preliminaries}
\subsection{The Model}
\label{subsec:TheModel}
\begin{figure}
\includegraphics[scale=1]{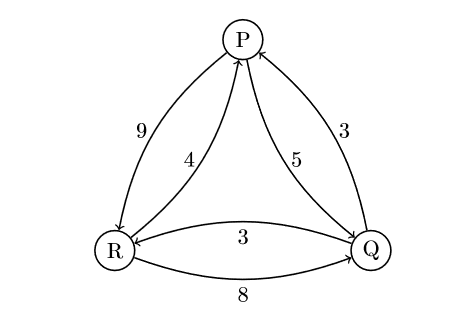}
\caption{Ordered Interaction Model}
\label{fig:InteractionGraph}
\end{figure}

In order for us to model the Tribler network in a graph structure we look into the number of different models defined in \cite{Sybil-resistant trust mechanisms in distributed systems} There, the concept of an ordered interaction model is introduced, which we will briefly recap.
\begin{definition}[Ordered Interaction Model]\ \\
Define an ordered interaction model as $M=\left\langle P,I,a\omega\right\rangle$, which is uniquely determined by the following 4 sets:
\begin{itemize}
\item $P:$ a finite set of agents.
\item $I$ a finite set of interactions.
\item $a:I\rightarrow{}P\times{}P$, a function mapping each interaction to the participants involved .
\item $\omega:I\times{}P\rightarrow{}\mathbb{R}_{\geq{}0}$, a function mapping an interaction and an agent to the contribution made by the agent in that interaction. \\
\end{itemize}
For any $p\not\in{}a(i)$ it must hold $\omega{}(i,p)=0$. Note that $\omega$ also allows for negative values, in case of leeching. For any agent the interactions involving this agent must be completely ordered temporally. 
\end{definition}
From this mathematical framework we now derive the graph structure on which we will instantiate the trust mechanism. 
\begin{definition}[Interaction Graph]
Given an ordered interaction model $M=\left\langle P,I,a,\omega\right\rangle$, we define an interaction graph $G = (\mathcal{V}, \mathcal{E})$ by the sets
\begin{itemize}
\item $\mathcal{V}=\left\lbrace v_p | p\in{}P\right\rbrace$
\item $\mathcal{E}=\left\lbrace (v_p,v_q) | \exists i\in I, a(i)=(p,q) \right\rbrace$ \vspace{1em}\\
\end{itemize}
Here, the network is given by a directed graph $(\mathcal{V}, \mathcal{E})$ with a set $\mathcal{V}$ of $N$ vertices or nodes and a set $\mathcal{E}$ of edges. The vertices correspond to the agents, or peers, in the network. The edges reflect interactions between agents.\vspace{1em}\\
The edge weights for an edge $(v_p,v_q)\in\mathcal{E}$ in this graph are determined by \vspace{0.3em}\\
\[ 
\omega((v_p, v_q)):=\sum_{\left\lbrace i\in I:a(i)=(p,q)\right\rbrace}\omega(i,p) \vspace{0.6em}\\
\]
\end{definition}
\noindent Edges are unidirectional and weighted, whereby the weight $\omega(a,b)$ of an edge connecting vertices $a$ and $b$ corresponds to the net data flow between two peers, i.e. the difference between up- and downloaded data. Two vertices are only ever connected by at most one directed edge. The direction of the edge is determined by the absolute value of the net data flow in between two nodes, i.e. if node $a$ has a surplus of uploaded data over node $b$ then the edge connecting the two will point towards $a$. With each transaction, edges are either added to the graph or modified in their weight. This means that, if for instance the absolute difference of up- and downloaded data changes, such that the node that had previously seeded more than leeched, now finds itself in "debt", then the direction of the edge is changed. An example of an interaction graph can be found in Figure \ref{fig:InteractionGraph}, with agents P,Q and R.

\subsection{Interaction Graphs in Tribler}
\label{subsec:InteractionGraphsInTribler}
As we have already discussed in \ref{subsec:Trustchain}, interactions in the Tribler network are recorded in a distibuted storage structure called the TrustChain. In the TrustChain, every block corresponds to an interaction. We can visualize the trustchain with what \cite{Sybil-resistant trust mechanisms in distributed systems} refers to as an ordered interaction graph, were every node corresponds to a block in the trustchain and the edges represent the hash pointers in the chain. However, in order for us to obtain a trust mechanism for agents in the Tribler network by running the pagerank algorithm, we need to transform our trustchain graph into an interaction graph. In order to do this, we take a closer look at the blocks contained in the tribler blockchain. \vspace{1em}\\

\noindent Blocks are tables made up of the following columns:
\begin{itemize}
\item type
\item tx
\item public key
\item sequence\_{}number
\item link\_{}public\_{}key
\item link\_{}sequence\_{}number
\item previous\_{}hash
\item signature
\item block\_{}timestamp
\item insert\_{}time
\item block\_{}hash
\end{itemize}

\noindent The tx column contains a dictionary of four keys: total\_{}up, total\_{}down, up and down. These values correspond to the data shared between the two nodes involved in the transaction. The total\_{}up and total\_{}down values represent the accumulated data flow in between the two nodes over the entire time. The public\_{}key value represents the public key of the requester of the transaction while the link\_{}sequence number is that of the responder. The sequence number is the numbering of the blocks in the chain of the requester and the link\_{}sequence\_{}number then represents the numbering of the block in the responder's chain. \vspace{1em}\\

\noindent In the interaction model, every node in the graph corresponds to a public key in the network. We retrieve all nodes in the graph from the public\_{}key and link\_{}public\_{}key fields of all trustchain blocks. We then acquire the edges from the up and down values of the "tx" dictionary. We aggregate these over all transactions involving the two agents and then determine the weight and subsequently the direction of the edge in between two nodes. This is repeated for all pairs of nodes until we have generated an entire interaction graph. We did this with the help of the Python sqlite3 module. 
\section{The Algorithm}
\label{sec:TheAlgorithm}
\subsection{PageRank}
\label{subsec:PageRank}
Traditionally, PageRank is computed through a method called the Power Iteration. The Power Iteration is determined in the following way. Let $n$ denote the number of nodes in the network. Define the $n\times{}n$ hyperlink matrix $P$ such that if $i$ is a node with $k$ outgoing edges with respective edge weights $\omega_{ij}$, for $k$ nodes $j$ connected to $i$ then $P_{ij}=\frac{\omega_{ij}}{\sum_{l=1}^{k}\omega_{il}}$ and $P_{ij}=0$ if $j$ is not connected to $i$ at all. The entries of P can then be interpreted as the probability of a random walk of the network hopping from node $i$ to node $j$, given that it has reached node $i$. If a page has no outgoing links at all, it is called a dangling node and the transition probability is then spread evenly among all nodes, i.e. $P_{ij}=\frac{1}{n}$.\vspace{1em}\\

\noindent This can be intrepreted as a random surfer on the graph hopping from node to node along the edges and then teleporting to a random node in the graph with some reset probability $c$ at every stop. It will also teleport once it reaches a dangling node.\vspace{1em}\\

\noindent Thus, the PageRank is defined as a stationary distribution of a Markov chain whose state space is the set of all nodes, and its transition matrix is given by 
\[ 
\tilde{P} = (1-c)P + c\left(\frac{1}{n}\right)E 
\]
where $E$ is a $n\times{}n$ matrix with all values equal to one and $c\in{}(0,1)$ is the reset probability. This matrix is stochastic, aperiodic and irreducible and therefore there exists a unique vector $\pi$ such that 
\[
\tilde{P}\pi = \pi,\quad \textrm{with } \pi\underline{1} = 1 
\]
This vector $\pi$ is then called the PageRank vector. The values $\pi_i$ can be interpreted as the probability of a random surfer landing on page $i$ in an infinite random walk of the network. It denotes the importance of nodes in the graph. In our case this importance to the network can also be interpreted as the nodes' trustworthiness. \vspace{1em}\\

\noindent The Power Iteration can be computed in an iterative manner. One begins with $\pi_0(i)=\frac{1}{n}$ for all nodes $i=1,...,n$. The values are then computed iteratively as follows. 
\[
\pi_{k+1}(i) = \frac{1-c}{n} + \sum_{\left\lbrace j\,|\,(j,i)\in\mathcal{E}\right\rbrace}\left(\frac{\pi_k(j)}{outdeg(j)}\right)c 
\]

\noindent This iteration is finished, once the difference of the consecutive values passes a certain treshold $\varepsilon$, or the number of iterations reaches a certain predetermined upper bound.\vspace{1em}\\

\noindent The PageRank values can be personalized to a particular seed node. In that case the pagerank values of all nodes in the network depend on the initial choice of the seed node. They are, so to speak, the pageranks determined from the perspective of the seed node. In that case the transition matrix is given by 
\[
\tilde{P} = (1-c)P + c\left(\frac{1}{n}\right)\tilde{E} 
\]
whereby $\tilde{E}$ is a matrix of only zeros, with a column of ones at the i-th place, with $i$ being determined by the seed node. \vspace{1em}\\

\noindent The iterative computation then follows a slightly different formula. For the pagerank of the seed node it is given by 
\[
\pi_{k+1}(i) = 1-c + \sum_{\left\lbrace j\,|\,(j,i)\in\mathcal{E}\right\rbrace}\left(\frac{\pi_k(j)}{outdeg(j)}\right)c 
\]
and for all other nodes 
\[
\pi_{k+1}(i) = \sum_{\left\lbrace j\,|\,(j,i)\in\mathcal{E}\right\rbrace}\left(\frac{\pi_k(j)}{outdeg(j)}\right)c 
\]
\subsection{Monte Carlo Methods for PageRank approximations}
\label{subsec:MonteCarloMethodsForPageRankApproximations}
In \cite{Monte Carlo methods in PageRank computation: When one iteration is sufficient} it is observed that the end-point of a random walk of the network that starts from a random page and can be terminated at each step with probability $1-c$, appears to be a sample from the distribution of $\pi$. Thus we find a random walk based algorithm for estimating the pagerank values of a network. By repeating the random walks of the graph many times, an estimate of $\pi_j$ for $j = 1,...,n$ can be given by the number of times a run crosses $j$, divided by the total number of nodes crossed by all random walks. This method for approximating PageRank is known as the Monte Carlo method. \vspace{1em}\\

\noindent The Monte Carlo method for estimating the PageRank vector has several advantages over the power iteration method. Let's recall the three requirements we defined for the trust algorithm in Problem Description. Personalization, locality and incrementality. The Monte Carlo algorithm satisfies all three of these requirements and we will show that it provides a good estimation of PageRank after relatively few iterations.\vspace{1em}\\

\begin{figure}
\includegraphics[scale=0.6]{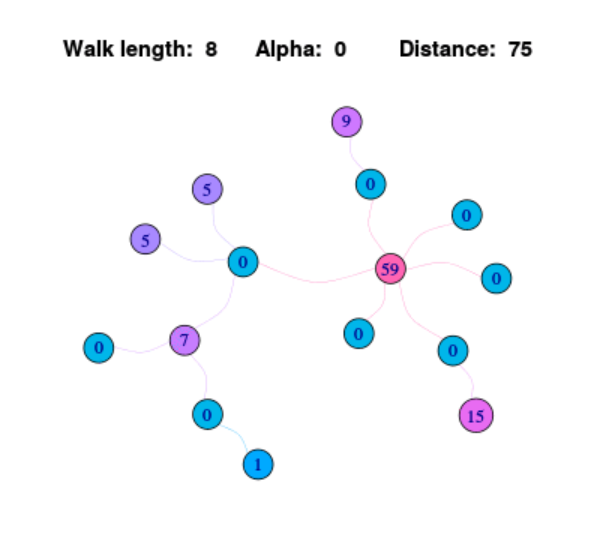}
\caption{Monte Carlo based PageRanks}
\label{fig:RandomWalks}
\end{figure}

\noindent Monte Carlo algorithms are motivated by the following formula
\[
\pi = \frac{1-c}{n}\underline{1}^T\left[\mathbbm{1} - cP\right]^{-1} = \frac{1-c}{n}\underline{1}^{T}\sum_{k=0}^{\infty}c^kP^k 
\]
\noindent which suggests a simple way of sampling from the PageRank distribution. In \cite{Monte Carlo methods in PageRank computation: When one iteration is sufficient} 5 different algorithms to estimate the personalized pagerank, motivated by the equation above, are introduced. We chose to implement the fourth algorithm. We run a simple random walk $(X_t)_{t\geq{}0}$, $m$ times starting at the seed node in the network that either stops with probability $c$ at every node reached, or when a dangling node is reached, else it transitions along the edges as determined by matrix $P$. The length of these random walks varies around an average of $\frac{1}{c}$. According to the equation above the nodes reached by these random walks then should approximate the distribution of $\pi$. Figure  \ref{fig:RandomWalks} is a visual representation of this particular Monte Carlo method.
Let $W_{ij}$ be a random variable denoting the number of times the random walk $(X_t)_{t\geq 0}$ reaches node $j$ given that it originated at node $i$. Formally,
\[ \mathbb{P}(W_{ij} = x) = \mathbb{P}\left(\left[\sum_{t=0}^{\infty}\mathbbm{1}_{\left\lbrace X_t=j\right\rbrace}\right]=x|X_0=i\right)
\]
Then, the estimator for $\pi$ obtained by this algorithm is given by
\[ 
\bar{\pi}_j = \left[\sum_{l=1}^{m}\sum_{i=1}^{n}W^{(l)}_{ij}\right]\left[\sum_{l=1}^{m}\sum_{i,j=1}^{n}W^{(l)}_{ij}\right]^{-1} 
\]
where $W^{(l)}$, $l\geq{}1$ are independent random variables of the same distribution as $W_{ij}$.\vspace{1em}\\

\noindent Theorem 1 in \cite{Monte Carlo methods in PageRank computation: When one iteration is sufficient} states that the estimator $\bar{\pi}_j$ converges to $\pi_j$ as the number of random walks goes to infinity. Thus $\bar{\pi}$ is a consistent estimator. Additionally, it is shown that this particular Monte Carlo method achieves an error of about 1\% for the highest 12 nodes. Thus, the defined $\bar{\pi}$'s are accurate approximations of the actual PageRank values $\pi$. In \cite{Fast incremental and personalized pagerank} it was shown that this estimator is, in fact unbiased and sharply concentrated around its expected value. We now evaluate this algorithm based on the three requirements listed in \ref{subsec:ProblemDescription}.
\begin{itemize}
\item[1] {\bf Personalization} \\ Recall the primary idea behind the PageRanks of nodes was for a node seeking to acquire data, to evaluate the peers in its vicinity. There is no central authority which it can query for these values and it cannot necessarily rely on other agents sharing information about their previous transactions with other nodes honestly. Therefore, the node must determine these values by itself. We adjust the Monte Carlo method such that all random walks originate from the same node, i.e. the downloader, also known as the seed node. The PageRanks are then computed analogously, but from the "perspective" of the seed node. The alternative PageRank vector obtained through this method is called personalized PageRank.
\item[2] {\bf Locality} \\ Recall the second requirement we had for this algorithm. Nodes do not query the entire network for data. They direct their focus primarily towards other nodes within their vicinity. This is due to transitive trust declining rapidly over an increasing number of hops through the network. The Monte Carlo algorithm enables exactly this, by applying random walks with relatively high reset probabilities, e.g. values in between 0.2 and 0.5 we find that the PageRanks are more focused around a node's immediate neighbourhood as opposed to weighting all nodes alike.
\item[3] {\bf Incremental Computation}\\ The tribler network does not stay still. Accounts are made as well as deleted and new transactions transpire continuously. Edges are added and removed or have their weight changed through transactions. Consequently the PageRank values fluctuate and have to be recomputed everytime the graph structure changes. It is simply unsustainable to recompute all random walks from scratch everytime the network is updated. Small alterations made to the graph, such as the addition or deletion of individual nodes and/or edges, are unlikely to impact a significant proportion of random walks in the network. Therefore it's just not necessary to recompute all random walks everytime the graph changes. \vspace{1em}\\
Only those random walks that pass through an edge or node that has been altered, need to be reviewed. And they only need to be recomputed starting at the last node they reached before they passed the modified region (edge or node). Let's say a node is removed from the network. All random walks that reach a node, the removed node was connected to, are then recomputed starting at the that node. If an edge is removed then all random walks that reach the source node of the removed edge are recomputed starting at the source node. The same goes for when an edge is added. This, of course is an alternative that is computationally far less expensive than a rerun of the entire algorithm everytime an edge is added or removed. \vspace{1em}\\
It should be noted here that it is far more common for new edges and nodes to arrive rather than existing edges and nodes to leave the network.\vspace{1em}\\
\end{itemize}
In \cite{Fast incremental and personalized pagerank} the following two theorems about the expected amount of additional update work for arriving and disappearing edges, were proved.

\begin{theorem}
Let $(u_t, v_t)$ be the $t^{th}$ random edge that is added to the graph and let $M_t$ be the random variable that determines the number of random walk segments that subsequently need to be updated. Finally, let $R$ denote the number of random walks and $l_i$ the length of the $i^{th}$ random walk. It then holds 
\[
\mathbb{E}[M_t] \leq \sum_{i=1}^{R}l_i\mathbb{E}[\frac{\pi_{u_t}}{outdeg_{u_t}(t)}] 
\]
Additionally, it holds 
\[
\mathbb{E}[\frac{\pi_{u_t}}{outdeg_{u_t}(t)}] = \frac{1}{t} 
\]
So it is
\[
\mathbb{E}[M_t] \leq \frac{1}{t}\sum_{i=1}^{R}l_i 
\]

\noindent For each random walk segment that needs to be updated, we compute a random walk from the corresponding source node. Hence, we can expect $\frac{1}{c}$ work per recomputed random walk. Now, we can sum up the expected amount of update work over $m$ edge arrivals and we obtain
\[
\mathbb{E}[\sum_{t=1}^{m}M_t] \leq \frac{1}{\varepsilon}\sum_{i=1}^{R}l_i \sum_{t=1}^{m}\frac{1}{t}
\]
Seeing as the harmonic series is bounded from above by the natural logarithm, we finally obtain
\[
\mathbb{E}[\sum_{t=1}^{m}M_t]\leq\frac{1}{\varepsilon}\sum_{i=1}^{R}l_i \ln{m} \approx \frac{R}{\varepsilon^2}
\]
\end{theorem}
For the counterpart of edges leaving the graph we have the following result.
\begin{theorem}
\noindent For a network with $m$ edges, if a randomly chosen edge leaves the graph, then the expected amount of work necessary to update the walk segments is at most
\[
\frac{1}{\varepsilon}\sum_{i=1}^{R}l_i\approx\frac{R}{\varepsilon^2}
\]
\end{theorem}
\noindent From this we learn that the cost to keep the PageRank approximations updated is only logarithmically larger than the cost of the initial computation and that the marginal update cost decreases with later edges making it increasingly cost effective over time, enabling real-time updates at a later stage.

\section{Implementation}
\label{sec:Implementation}
We wrote a Python class implementing the algorithm. The code can be found in \url{https://github.com/alexander-stannat/Incremental-Pagerank/blob/master/Page\_{}Rank2.py}
\subsection{Unit Testing}
\label{subsec:UnitTesting}
In order to determine whether our algorithm runs correctly, we ran a set of 5 different unit tests. In the last one, we generated a random graph with number of nodes in between 2 and 10000 and 2 edges per node. We then computed the Monte Carlo pageranks as well as the results of the power iteration. We compare the results and assert that the two vectors do not diverge by more than $10\%$. Thereafter, we update the graph by randomly adding and removing edges and nodes, and then update the values of the Monte Carlo PageRanks incrementally, as discussed in \ref{subsec:ProblemDescription}. We also recompute the values of the power iteration and compare the two values again. We find that the values never diverge by more than $10\%$ for any of the randomly generated graphs. 
\subsection{Convergence Testing}
\label{subsec:ConvergenceTesting}
\begin{figure}
\includegraphics[scale=0.3]{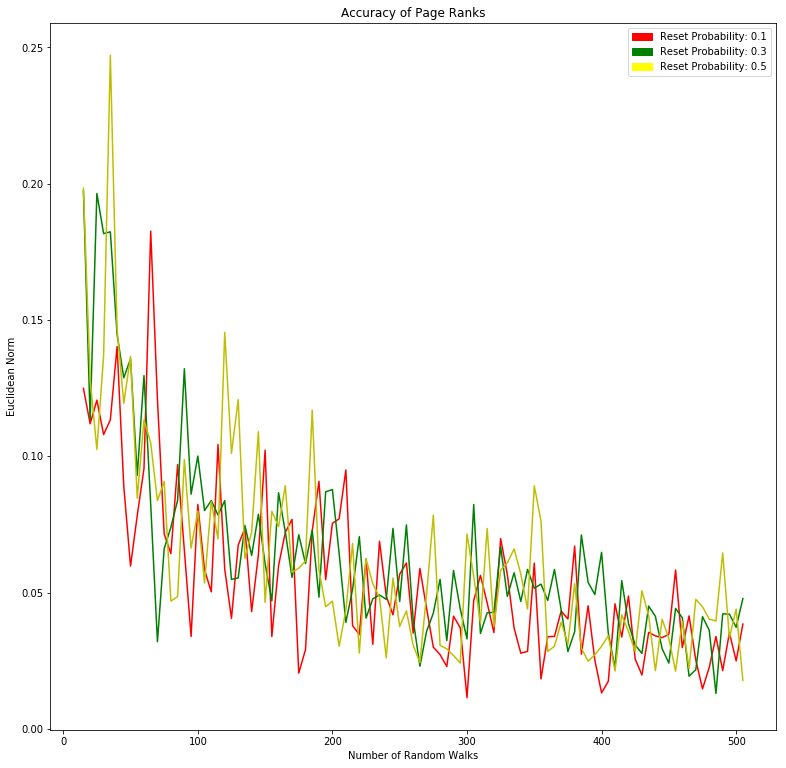}
\caption{Accuracy of the Monte Carlo PageRanks by number of random walks}
\label{fig:Accuracy1}
\end{figure}

Next we determine the rate at which the Monte Carlo algorithm converges to its final values for different sets of graphs and input parameters. Again, we randomly generate a graph, with the number of nodes ranging from 2 to 100. We begin with the random generation of the graph. This works analogously to our previous example above in \ref{subsec:UnitTesting}. We determine the values of the Monte Carlo pagerank for reset probabilities $0.1, 0.3$ and $0.5$. We plot the accuracy of the method relative to the power iteration against the number of random walks ranging from 10 to 500. We obtain the following results given in Figure \ref{fig:Accuracy1}. \vspace{1em}\\

\noindent Figure \ref{fig:Accuracy1} depicts the accuracy of our Monte Carlo computation of the pageranks relative to the power iteration method. We can clearly see a downward trend in the error values over the number of random walks. It is pretty obvious that the Monte Carlo PageRanks become more accurate as the number of random walks through which they are computed, increases. We see that the error values for small numbers of random walks can be as high as $20\%$ and go down to under $5\%$ which are reasonably accurate values as discussed in \ref{subsec:UnitTesting}. We also see that the reset probability has no effect on the accuracy, which is also expected. One should remark the high variance, we can find in these values, as seen in the jaggedness of the curves. This is due to the probabilistic nature of the algorithm, i.e. the random walks. This means the accuracy can vary quite significantly for any number of random walks and any reset probability. \vspace{1em}\\

\begin{figure}
\includegraphics[scale=0.385]{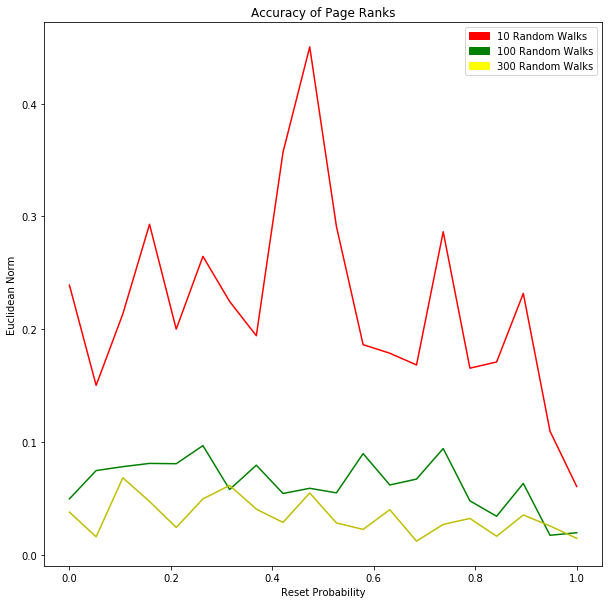}
\caption{Accuracy of the Monte Carlo PageRanks by reset probabilities}
\label{fig:Accuracy2}
\end{figure}

\noindent Next, we determine whether the reset probability of the random walks has any impact on the accuracy of the algorithm. We plot three error values of the Monte Carlo method for different numbers of random walks $(10, 100, 300)$ and obtain the values in Figure \ref{fig:Accuracy2}. \vspace{1em}\\ 

\noindent We see in Figure \ref{fig:Accuracy2} that the accuracy of the Monte Carlo PageRanks does have a slight dependence on the reset probability as well as the number of random walks. For reset probabilites that are further away from the bounds, 1 and 0, we find that the error values tend to be larger. This makes sense as well. Large reset probabilities generate short random walks, which are less volatile than long ones. Therefore there is less room for probabilistic fluctation, making them more accurate. Analogously, small reset probabilites generate very long random walks that don't reset at all, unless they reach a dangling node. This means the visit times of these random walks are more stable because they cover the entire graph, and therefore we obtain a smaller error. It should also be added that this effect is weakened as the number of random walks increases.
\subsection{Sybil Resistance}
\label{subsec:SybilResistance}
\begin{figure}
\includegraphics[scale=0.3]{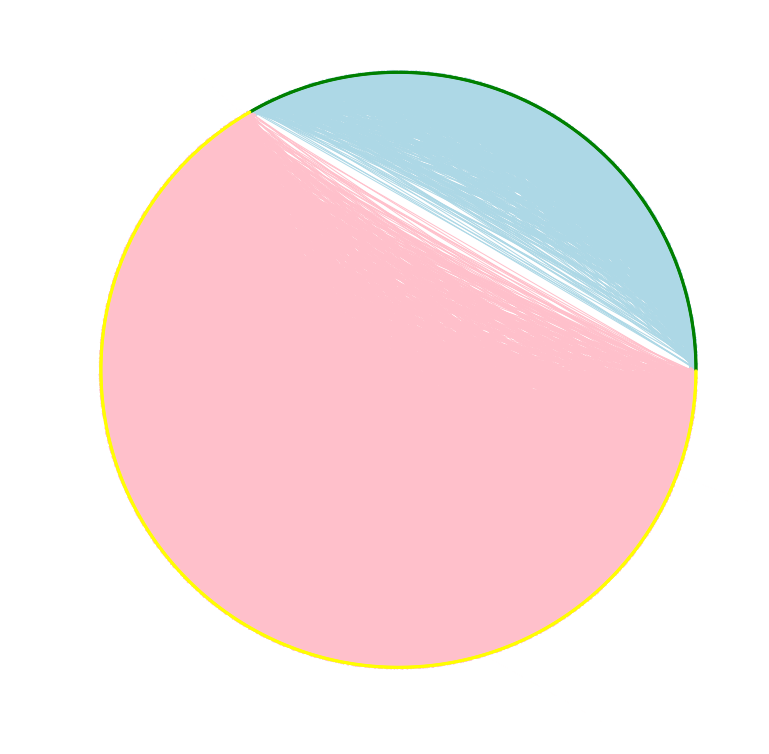}
\caption{Image of a network split into an honest region and a sybil region}
\label{fig:SybilRegion}
\end{figure}

In this section we discuss the properties of sybil resistance of the Monte Carlo method. We randomly generate a graph for an honest region, analogously to the method discussed in \ref{subsec:UnitTesting}. The honest region consists of 500 random nodes and 2000 randomly generated interconnecting edges each weighted from 0 to 10. We also generate a sybil region made up of 1000 nodes which is even more densely connected than the honest region, with 10 edges per node. In Figure \ref{fig:SybilRegion} the honest region is given by the green nodes and light blue edges, while the sybil region has yellow nodes and pink edges. The seed is node 0 and lies in the honest region.\vspace{1em}\\

\noindent Next, we add attack edges connecting the honest and the sybil region in the graph. We compute the Monte Carlo PageRanks of the given graph for every number of attack edges in between 0 and 500. There are several attack edges that are directly connected to the seed node 0. The Monte Carlo algorithm is run with 200 random walks and several different reset probabilities. For each of those PageRank vectors we create a list "ordered\_{}nodes", which orders the nodes in the graph according to their pageranks in a descending order. Using this ordered vector, we determine the ROC curve and the area under the ROC curve as a measure of sybil resistance of the Monte Carlo method. \vspace{1em}\\

\begin{figure}
\includegraphics[scale=0.3]{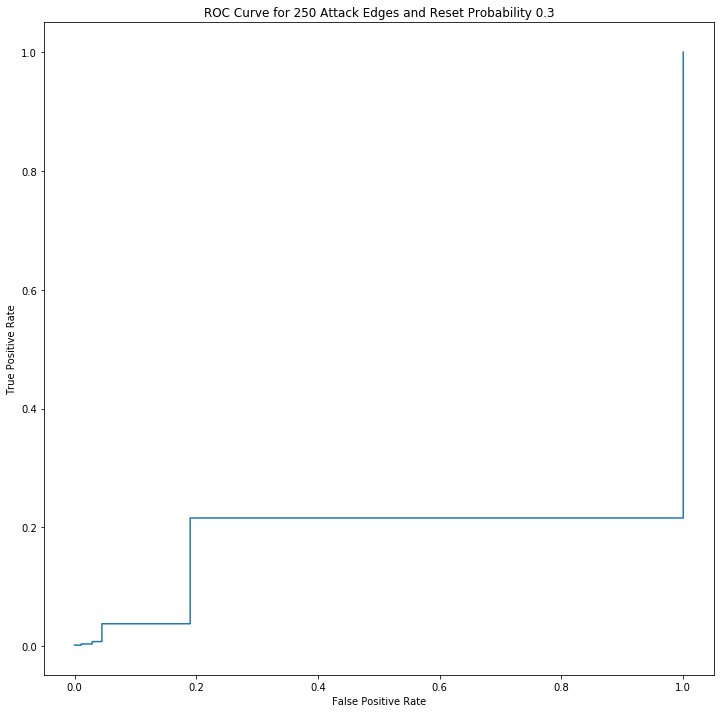}
\caption{ROC curve for 0 attack edges and reset probability 0.3}
\label{fig:ROCCurve1}
\end{figure}

\begin{figure}
\includegraphics[scale=0.3]{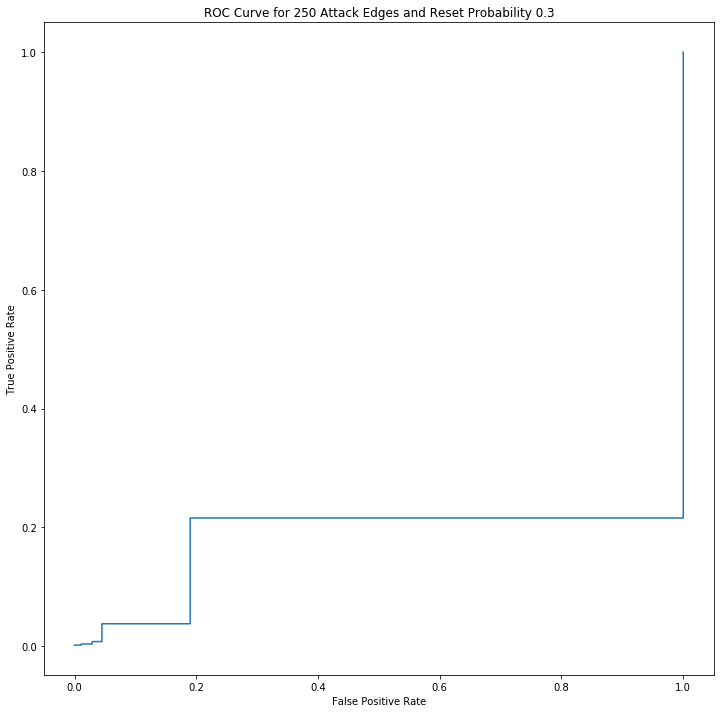}
\caption{ROC curve for 250 attack edges and reset probability 0.3}
\label{fig:ROCCurve2}
\end{figure}

\begin{figure}
\includegraphics[scale=0.3]{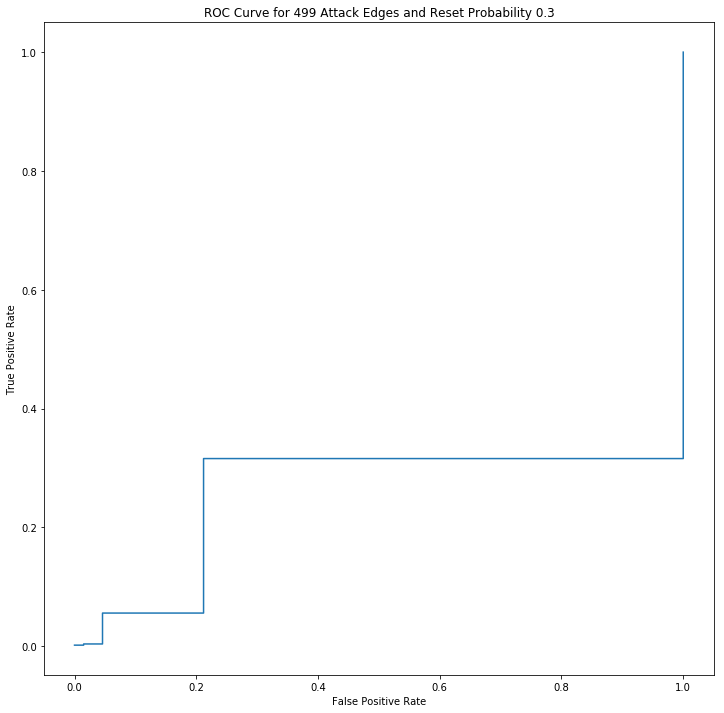}
\caption{ROC curve for 499 attack edges and reset probability 0.3}
\label{fig:ROCCurve3}
\end{figure}

\noindent ROC curves are used to show the connection between sensitivity and specificity for every possible cut-off for diagnostic tests. In addition, the area under the ROC curve gives an idea about the benefit of using the test in question. In our case, it is determined as follows. We let the cut-off point run from the first to the final node. For each, we label the nodes beneath it as sybils and the nodes above as honest. The abscissa of the ROC curve is then given by the false positive rate for each cut-off point and is plotted against the ordinate given by the corresponding true positive rate. We determine the ROC curve for each number of attack edges and a set of different reset probabilities (0.1, 0.3, 0.5, 0.7). We obtain the plots given in Figures \ref{fig:ROCCurve1}, \ref{fig:ROCCurve2}, \ref{fig:ROCCurve3}.\vspace{1em}\\

\begin{figure}
\includegraphics[scale=0.6]{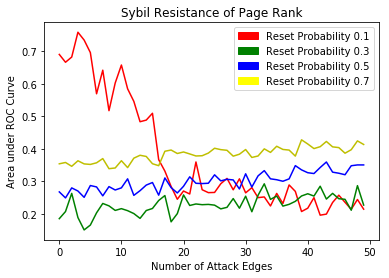}
\caption{Area under the ROC curve by number of attack edges for various reset probabilities}
\label{fig:ROCArea1}
\end{figure}

\begin{figure}
\includegraphics[scale=0.6]{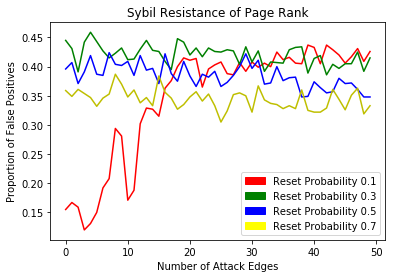}
\caption{Proportion of false positives by number of attack edges for various reset probabilites}
\label{fig:ROCArea2}
\end{figure}

\begin{figure}
\includegraphics[scale=0.6]{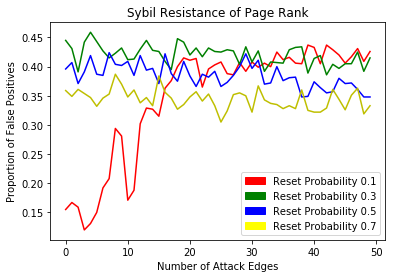}
\caption{Proportion of false negatives by number of attack edges for various reset probabilities}
\label{fig:ROCArea3}
\end{figure}

\noindent Next, we determine the area under the ROC curve for every ROC curve computed. The area under the ROC curve (AUROC) of a test can be used as a criterion to measure the test's discriminative ability. We obtain the graphs in Figures \ref{fig:ROCArea1}, \ref{fig:ROCArea2}, \ref{fig:ROCArea3}.\vspace{1em}\\

\noindent We can see in the plots above that the area under the ROC curve decreases monotonely for random walks with reset probability 0.1. This does not hold for shorter random walks. When we think about this, we find that it makes perfect sense. Given the size of the graph and the length of the random walks, most nodes in the honest network are never visited by any random walk. Hence they have a pagerank of 0. This means they are at the bottom of the ordered\_{}nodes vector and therefore labelled as sybil for almost all cut-off points, i.e. we have a very large proportion of false negatives. This does not hold for the random walks with reset probability of 0.1, because those random walks travers the entire graph. This leads to a slightly misleading conclusion. \vspace{1em}\\

\noindent Overall, we get a false positives rate of in between 0.35 and 0.45. The shorter the random walks, the lower this rate. We also find that for short random walks this rate actually decreases with the number of attack edges. This is because a larger number of attack edges enables random walks that have entered the Sybil region, to also escape it. \vspace{1em}\\

\begin{figure}
\includegraphics[scale=0.6]{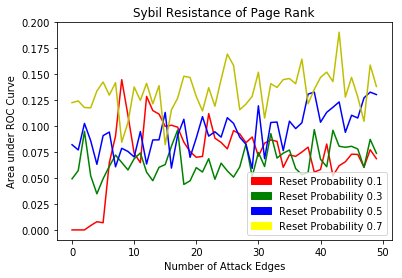}
\caption{Area under the ROC curve with nodes of PageRank 0 removed}
\label{fig:ROCArea4}
\end{figure}

\begin{figure}
\includegraphics[scale=0.6]{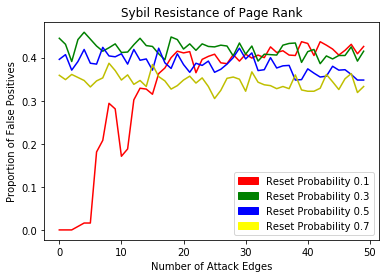}
\caption{Proportion of false positives with nodes of PageRank 0 removed}
\label{fig:ROCArea5}
\end{figure}

\begin{figure}
\includegraphics[scale=0.6]{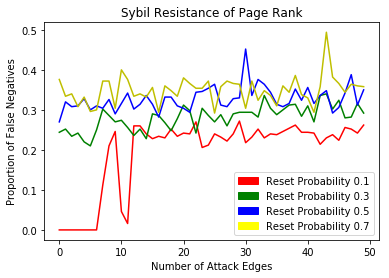}
\caption{Proportion of false negatives with nodes of PageRank 0 removed}
\label{fig:ROCArea6}
\end{figure}

\noindent In order to obtain results that are more representative of the method's sybil resistance, we remove all nodes with PageRanks of exactly 0 from the ordered\_{}nodes vector. With this, we obtain the results given in Figures \ref{fig:ROCArea4}, \ref{fig:ROCArea5}, \ref{fig:ROCArea6}. \vspace{1em}\\

\begin{itemize}
\item {\bf Area under the ROC Curve} \\
We find that the area under the ROC curve increases quite significantly after the 5th attack edge is added. This is because the 5th attack edge is the first attack edge that is directly connected to the seed node. Therafter it decreases moderately with the number of attack edges. For the remaining reset probabilities this does not seem to be the case. Instead the area under the ROC curve increases with the number of attack edges connecting the sybil and the honest region. This is due to the reasons we discussed above.

\item {\bf False Positives} \\
False positives are sybil nodes that are considered honest according to their pageranks. We find that for the reset probability of 0.1 the number of false positives increases abruptly after the 5th attack edge is added, again for the same reasons as mentioned above. The remaining reset probabilities exhibit rather constant numbers of false positives. The overall average of false positives for the algorithm in this network lies at around $35\%$.

\item {\bf False Negatives} \\
False negatives are honest nodes that are marked as sybils by the algorithm. We find again that the monte carlo algorithm with a reset probability of 0.1 exhibits a jump in false negatives at around 5 attack edges. The remaining PageRank values seem to be quite constant with a minor upward trend with the number of attack edges added. The overall average here is at around $30\%$.
\end{itemize}

\noindent We find that these values are not really satisfactory and that therefore the Monte Carlo PageRank algorithm is not practically sybil resistant. This makes sense to us. Random walk based PageRank algorithms are rarely sybil resistant, simply due to the random walks that they are based on. It is easy for a random walk to end up in the sybil region and there is no mechanism to prevent this in our current implementation. In the future we may want to implement a mechanism to enhance our algorithms sybil proofness, for instance by identifying attack edges. 
\section{Application to the Tribler Network}
\label{sec:ApplicationToTheTriblerNetwork}
We now apply our implementation of the Monte Carlo algorithm to the Tribler network.
\subsection{Running the Algorithm}
\label{subsec:RunningTheAlgorithm}
First we generate a graph of the Tribler network. We generate the graph as discussed in \ref{subsec:MonteCarloMethodsForPageRankApproximations}. There are 667124 blocks in our data set. This corresponds to a graph of 289 nodes. We first create our graph for all but 5 nodes and all but 20 edges. Then we compute the corresponding PageRanks. We update the graph by adding the missing nodes and edges to the graph. Then we update our random walks and recompute the PageRank values. In both cases we compare our values with the Power Iteration values and time the entire process. We employ our Monte Carlo algorithm with 300 random walks and a reset probability of 0.3. The algorithm has a runtime of 14.1710000038 seconds and we obtain following error values: 0.3978275442360459 for the euclidean norm and 0.19806694712219172 for the supremum norm. This is not quite as accurate as our previous results for smaller graphs and we conclude that there must be some scalability issues in the algorithm.

\subsection{Incremental Update}
\label{subsec:IncrementalUpdates}
Now, we update our graph by adding the missing edges and nodes. We recompute the Power Iteration PageRanks as before, but recompute the Monte Carlo PageRanks not from scratch, but by updating the necessary random walks, as discussed in \ref{subsec:MonteCarloMethodsForPageRankApproximations}. The algorithm has a runtime of 5.17000007629 seconds and we obtain the error values of 0.46419964703781746 for the euclidean norm and 0.19798405911144684 for the supremum norm. These values are slightly larger than the ones we obtained from the initial computation, but still reasonable. 

\section{Conclusion}
\label{sec:Conclusion}
In this project, we implemented an incremental personalized PageRank algorithm for ranking nodes in a peer-to-peer file sharing network, called Tribler using the Monte Carlo method. We chose algorithm 4 from \cite{Monte Carlo methods in PageRank computation: When one iteration is sufficient} with a set of standard random walks on a directed graph with weighted edges. The graph of the tribler network was generated by "flattening" the blockchain's blocks into a unidirectional graph with edge weights corresponding to the net data flow in between nodes.\vspace{1em}\\

\noindent We found that for relatively small graphs, the algorithm is quite fast (14 seconds for the initial computation and 5 seconds for the incremental updates). We find however that the error values tend to increase with the size of the network and for the initial computation of the pageranks of the tribler network we obtain an error of almost $40\%$ and $47\%$ for the incremental updates. One should note here that the application of the PageRanks is primarily to rank nodes based on their trustworthiness in order to select the most trusted nodes to interact with. This means, the PageRanks of lower-ranked nodes are not as relevant as the PageRanks of the highly-ranked nodes. An agent interested in a transaction will generate their trust ranking and most likely choose a peer among the top 5 ranked nodes. This means that some of the inaccuracy is negligible so long as the ranking of nodes at the top of the ranking is approximately correct and in the right order. Therefore the algorithm is still applicable and returns useful results for the Tribler application, despite the high error values. \vspace{1em}\\

\noindent We inspected the rate of convergence of the algorithm and found that the accuracy of values, relative to the power iteration's values improves exponentially by the number of random walks and shrinks down to values close to $3\%$ for a network consisting of 15 nodes. As the network grows these values increase as well. This is due to small inaccuracies accumulating by the iterations and thereby growing to ever larger values. We suspect that choosing the networkx library's approximation of the power iteration based PageRanks may not have been the best relative reference for our Monte Carlo PageRank. An example illustrates the possible inaccuracy of the Power Iteration's values. If we choose a network with a completely isolated seed node then we expect a PageRank vector consisting only of zeros and a single value of one. However, the networkx power iteration returns a vector with many very small values for most nodes other than the seed node. This perfectly illustrates where the large error values we obtained originated from. The networkx power iteration PageRanks have an innate inaccuracy, which accumulates and increases as the number of edges multiplies and the graph grows. \vspace{1em}\\

\noindent Finally, we tested the algorithm for its sybil resistance. We generated an honest region and a sybil region connected by a varying number of attack edges. We found that for an honest region with 500 nodes and a sybil region of 1000 nodes and a number of attack edges increasing up to 500 edges the algorithm exhibited a rather large percentage of false positives and negatives and a relatively low ROC curve. This was to be expected, simply due to the random walk based nature of the algorithm. In future work we may look into optimizing this algorithm to account for sybil attacks.

\section{Outlook}
Within the scope of this project we deliberately neglected a number of concerns and possible additions to the trust algorithm, which we list here now. In the coming weeks and months we will tackle some of these issues.
\begin{itemize}
\item {\bf Sybil Resistance} \\ We have seen that our algorithm is quite susceptible to sybil attacks, especially to those in which Sybil regions manage to create attack edges connecting directly to the seed node. Currently, we see the issue lie in the fact that it is too easy for Sybils to gain trustworthiness throught the transitivity of trust in the network. In the future we might want to look into making some additions to our current implementation in order to increase our algorithms sybil resistance.\vspace{1em}\\

\item {\bf CPU Hogging} \\ Running the Monte Carlo algorithm 5 seconds straight, everytime we incrementally update our values, occupies the CPU for too long periods at a time. Our current system uses a single core $100\%$ for this time span everytime it updates the trust values. These batch computations completely starve any other processes running on a user's CPU. This is problematic and should be replaced by an alterative system load profile, replacing the batch computations with continuous updates of the trust values of all agents in the network.\vspace{1em}\\

\item {\bf Accumulation of Errors} \\ We have seen in \ref{subsec:ConvergenceTesting} that the algorithm's first trust ranking entails a rather high error value. As the network continues to evolve, edges and nodes are added and removed and the trust ranking is updated incrementally in batch computations. Every update of the PageRank vector adds a new error to the already existing one and error values accumulate over time. In order to prevent errors from getting too large, the PageRank must be recomputed from scratch every so often. Before deploying this algorithm in the Tribler network it needs to be determined when and how regularly such resets must take place.\vspace{1em}\\

\item {\bf Relay Nodes} \\ The Tribler application is based on the Tor overlay network, in order to facilitate anonymous up- and downloading. This means that data transferred between two parties in the network is actually rerouted through many other nodes, acting as encryption layers. The current implementation of the algorithm treats the action of relaying equivalently to downloading and then uploading the same amount of data, oftentimes leading to an increase in trustworthiness. This should not be the case. Relaying {\it should} be treated separately from regular seeding and leeching and therefore also differently evaluated. 
\end{itemize}

\end{document}